\documentclass[lettersize,journal]{IEEEtran}
\usepackage{amsmath,amsfonts}
\usepackage{algorithmic}
\usepackage{algorithm}
\usepackage{array}
\usepackage[caption=false,font=normalsize,labelfont=sf,textfont=sf]{subfig}
\usepackage{textcomp}
\usepackage{stfloats}
\usepackage{url}
\usepackage{verbatim}
\usepackage{graphicx}
\usepackage{cite}
\usepackage{soul}
\usepackage{booktabs} 
\usepackage[table,xcdraw]{xcolor}
\usepackage[normalem]{ulem}
\usepackage{multirow}
\usepackage{balance}
\usepackage{tablefootnote}
\usepackage[utf8]{inputenc}
\definecolor{teal}{RGB}{62, 153, 159}          
\definecolor{lightred}{RGB}{255, 153, 153}    
\definecolor{grassgreen}{RGB}{157, 217, 124}     
\definecolor{darkgrayblue}{RGB}{150, 161, 171}  
\definecolor{skyblue}{RGB}{135, 206, 235}     
\definecolor{darkpurple}{RGB}{183, 137, 229}     
\begin{document}

The current version is `Preprint'.

This work has been submitted to the IEEE for possible publication. Copyright may be transferred without notice, after which this version may no longer be accessible.

This information aligns with the guidelines available at:

https://journals.ieeeauthorcenter.ieee.org/become-an-ieee-journal-author/publishing-ethics/guidelines-and-policies/post-publication-policies/
\newpage

\title{Nes2Net: A Lightweight Nested Architecture for Foundation Model Driven Speech Anti-spoofing}

\author{Tianchi Liu,~\IEEEmembership{Student Member}, Duc-Tuan Truong,~\IEEEmembership{Student Member}, Rohan Kumar Das,~\IEEEmembership{Senior Member}, \\ Kong Aik Lee,~\IEEEmembership{Senior Member}, Haizhou Li,~\IEEEmembership{Fellow}
\thanks{Tianchi Liu and Haizhou Li are with the Department of Electrical and Computer Engineering, National University of Singapore, Singapore. Tianchi Liu is also with LIGHTSPEED, Singapore (email: tianchi\_liu@u.nus.edu);}
\thanks{Duc-Tuan Truong is with the Nanyang Technological University, Singapore (email: truongdu001@e.ntu.edu.sg);}
\thanks{Rohan Kumar Das is with the Fortemedia Singapore, Singapore (email: ecerohan@gmail.com);}
\thanks{Kong Aik Lee is with the Department of Electrical and Electronic Engineering and the Research Centre for Data Science \& Artificial Intelligence, The Hong Kong Polytechnic University, Hong Kong (e-mail: kong-aik.lee@polyu.edu.hk);}
\thanks{Haizhou Li is also with the Shenzhen Research Institute of Big Data, School of Artificial Intelligence, School of Data Science, The Chinese University of Hong Kong, Shenzhen, China (email: haizhouli@cuhk.edu.cn).}
}
\markboth{IEEE Transactions on Information Forensics and Security}%
{Shell \MakeLowercase{\textit{et al.}}: A Sample Article Using IEEEtran.cls for IEEE Journals}


\maketitle

\begin{abstract}
Speech foundation models have significantly advanced various speech-related tasks by providing exceptional representation capabilities. However, their high-dimensional output features often create a mismatch with downstream task models, which typically require lower-dimensional inputs. A common solution is to apply a dimensionality reduction (DR) layer, but this approach increases parameter overhead, computational costs, and risks losing valuable information.
To address these issues, we propose Nested Res2Net (Nes2Net), a lightweight back-end architecture designed to directly process high-dimensional features without DR layers. The nested structure enhances multi-scale feature extraction, improves feature interaction, and preserves high-dimensional information. 
We first validate Nes2Net on CtrSVDD, a singing voice deepfake detection dataset, and report a 22\% performance improvement and an 87\% back-end computational cost reduction over the state-of-the-art baseline. Additionally, extensive testing across four diverse datasets: ASVspoof 2021, ASVspoof 5, PartialSpoof, and In-the-Wild, covering fully spoofed speech, adversarial attacks, partial spoofing, and real-world scenarios, consistently highlights Nes2Net’s superior robustness and generalization capabilities. The code package and pre-trained models are available at \url{https://github.com/Liu-Tianchi/Nes2Net}.

\end{abstract}


\section{Introduction}
\IEEEPARstart{S}{peech} foundation models, such as wav2vec 2.0~\cite{wav2vec2}, HuBERT~\cite{HuBERT}, and WavLM~\cite{WavLM}, have revolutionized speech processing by leveraging large-scale pretraining to capture complex acoustic and linguistic patterns~\cite{TERA}. This has driven notable advances in automatic speech recognition (ASR)~\cite{9801640}, speaker verification (SV)~\cite{ESPnet}, and other speech applications.

Beyond traditional tasks, speech foundation models also show great promise in addressing critical security concerns, particularly speech anti-spoofing (also referred to as deepfake detection)~\cite{10.1145/3714458}. With the growing sophistication of spoofing techniques, such as voice conversion, ensuring the reliability and security of speech-driven systems has become a pressing concern~\cite{10650962, 10830534, das20c_interspeech, SpoofCeleb, du2025codecfake}. Leveraging the rich representations of these foundation models could significantly improve the robustness and generalization of anti-spoofing systems~\cite{chen24o_interspeech, SLIM, hemlata_wav2vec2}. 

While speech foundation models offer exceptional representations, their high-dimensional feature outputs present significant challenges for downstream tasks. Downstream models used in tasks like speech anti-spoofing typically require lower-dimensional features~\cite{hemlata_wav2vec2, SVDD_i2r, Mamba}. To address this mismatch, a common approach is to introduce a dimensionality reduction (DR) layer, usually implemented as a fully connected (FC) layer for transforming high-dimensional features into lower-dimensional features. However, this conventional strategy presents notable drawbacks. 
Given that downstream classifiers are typically compact~\cite{hemlata_wav2vec2, SVDD_i2r}, the DR layer alone often consumes a substantial portion of the parameters and computational resources within the entire back-end model. Moreover, directly projecting high-dimensional features in a one-shot manner through an FC layer leads to the loss of important information, reducing the effectiveness of speech foundation models. These issues highlight the need for a more efficient and effective solution to bridge the dimensionality gap and fully utilize speech foundation models in downstream tasks.

To address these challenges, we propose Nested Res2Net (Nes2Net) to process high-dimensional features from speech foundation models, eliminating the need for a DR layer while preserving the richness of the original representations. By addressing key limitations of DR layers, such as excessive computational cost and information loss, Nes2Net offers a more efficient and effective solution. This design makes it particularly suitable for tasks requiring a balance of high performance and efficiency, such as speech anti-spoofing. The key contributions of this work can be summarized as follows:

\IEEEpubidadjcol

\begin{itemize}
    \item \textbf{Novel Architecture}: We introduce Nes2Net, a new approach that effectively addresses the limitations of DR layers. Nes2Net retains the expressive power of high-dimensional features while reducing model complexity.
    \item \textbf{Enhanced Performance, Efficiency, and Generalization}: Our method demonstrates a 22\% performance gain and an 87\% reduction in computational costs compared to the state-of-the-art baselines on the CtrSVDD dataset. Further experiments conducted on four additional datasets across various scenarios demonstrate strong generalization capability and consistently superior performance.
    \item \textbf{Reproducibility}: To facilitate further research and application, we make our scripts and pre-trained models publicly available.
\end{itemize}

\section{Related Work}

\subsection{Res2Net}

\begin{figure*}[t]
\includegraphics[width=\textwidth]{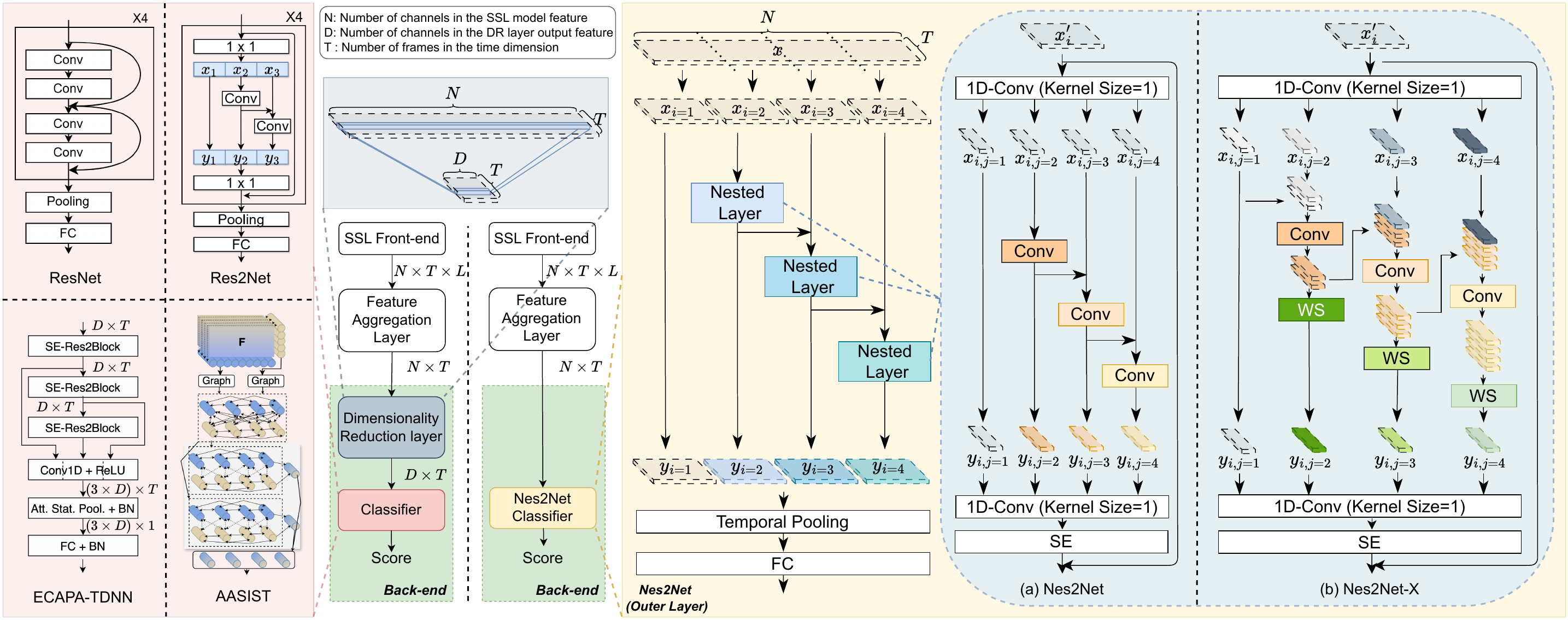}
\vspace{-0.2 in}
\caption{The block diagram of the speech foundation model-based speech anti-spoofing system, showcasing both the traditional back-end models and the proposed Nes2Net back-end. The traditional back-end models include a DR layer and a classifier, such as ResNet~\cite{resnet}, Res2Net~\cite{Res2Net}, ECAPA-TDNN~\cite{ECAPA_TDNN}, and AASIST~\cite{AASIST}. In contrast, the proposed Nes2Net back-end model features a DR layer-free design. Additionally, an enhanced version of its nested layer, named Nes2Net-X, is introduced to further improve performance.
Abbreviations used in the figure include: `FC’ (fully connected layer), `Conv’ (convolutional layer), `WS’ (weighted sum), `SE’ (squeeze-and-excitation module)~\cite{SE}, and `Att. Stat. Pool.’ (attentive statistics pooling)~\cite{Okabe2018}.}
\label{fig_Nes2Net}
\vspace{-0.1 in}
\end{figure*}

Res2Net~\cite{Res2Net} is a well-known architecture designed to extract multi-scale features. Unlike ResNet~\cite{resnet}, Res2Net uses hierarchical residual connections within a single block, allowing it to capture patterns across varying receptive fields simultaneously~\cite{Res2Net}.
This design offers proven advantages in speech-related tasks, such as SV~\cite{ERes2Net,ERes2NetV2,Golden_Gemini} and anti-spoofing~\cite{li21o_interspeech, 10096672, liu2024towards}, where capturing subtle variations and complex acoustic patterns is important.
{As shown in Fig.~\ref{fig_Nes2Net}, Res2Net (highlighted using a light red block) can also serve as a classifier within a speech foundation model-based anti-spoofing system.
Its ability to extract multi-scale features has led to superior performance over conventional models and motivates the design of Nested Res2Net in this work.

\subsection{{Hand-crafted Feature-based} Speech Anti-Spoofing Models}
\label{subsec_traditionalCM}
{Hand-crafted acoustic features (such as MFCC) are common choices for many earlier speech anti-spoofing systems. These systems} have evolved to effectively detect speech deepfakes~\cite{AASIST, RawBMamba}. For instance, the Channel-wise Gated Res2Net (CG-Res2Net)~\cite{li21o_interspeech} introduces a gating mechanism within the Res2Net architecture, enabling dynamic selection of channel-wise features to enhance generalization to unseen attacks. A widely recognized model is AASIST~\cite{AASIST}, which employs spectro-temporal graph attention layers to capture both temporal and spectral artifacts, thereby achieving efficient and accurate detection.
Given AASIST’s SOTA performance and its wide adoption in recent anti-spoofing challenges~\cite{SVDD_i2r, chen2024ustc}, we consider it as our main baseline for evaluation.

\subsection{Speech Foundation Models}
Speech foundation models are often referred to as Self-Supervised Learning (SSL) models due to their typical pre-training on large amounts of unlabeled speech data using self-supervised learning techniques. Examples include wav2vec 2.0~\cite{wav2vec2}, HuBERT~\cite{HuBERT}, and WavLM~\cite{WavLM}.
Unlike {hand-crafted acoustic features, which are limited in their ability to adapt to diverse and complex conditions}, self-supervised learning (SSL) models learn rich and generalized speech representations that can be effectively adapted to various downstream applications. This allows them to achieve superior performance in speech-related tasks, including speech anti-spoofing.

\subsection{Speech Foundation Model-based Anti-spoofing}
As discussed in the previous subsection, speech foundation models can capture more informative representations than handcrafted or raw acoustic features~\cite{WavLM}. This makes them highly effective for speech anti-spoofing, as they generalize well across datasets and are more robust to unseen attacks~\cite{hemlata_wav2vec2}. As a result, many recent anti-spoofing systems increasingly adopt these models as front-ends, feeding their features to the back-end classifiers and consistently outperforming traditional models~\cite{10890293, SVDD_i2r, 10888985}.

To connect these powerful front-end models to downstream classifiers, a feature aggregation layer is introduced, as shown in Fig.~\ref{fig_Nes2Net}.
This layer combines features from different SSL layers using methods such as a simple \textit{weighted sum} or attention-based methods like Squeeze-and-Excitation Aggregation (SEA)~\cite{SVDD_i2r} and Attentive Merging (AttM)~\cite{attentive_merge}.

Following the aggregation layer, the resulting features are passed to the back-end classifier, as shown in the green box of Fig.~\ref{fig_Nes2Net}. 
Existing methods typically use a DR layer, which reduces the high-dimensional features of $N$ channels (commonly $N = 1024$~\cite{wav2vec2, WavLM,MERaLiON}) to a lower dimension $D$ (e.g., $D = 128$~\cite{hemlata_wav2vec2,SVDD_i2r} or $D = 144$~\cite{Mamba,TCM}) to match the classifier’s input requirements.}
The classifier model then extracts features from the DR layer outputs and produces the final score. As illustrated in the red box of Fig.~\ref{fig_Nes2Net}, commonly used classifier structures include traditional models such as ResNet~\cite{resnet}, Res2Net~\cite{Res2Net}, ECAPA-TDNN~\cite{ECAPA_TDNN}, and AASIST~\cite{AASIST}.

The strong performance of these systems stems from their ability to capture rich speech representations, enabling more accurate distinction between real and spoofed speech. As a result, these systems have achieved SOTA results~\cite{SLS,TCM,10888281}, especially in recent challenges like ASVspoof 5~\cite{chen2024ustc, ASVspoof5_report}, CtrSVDD~\cite{SVDD_report,SVDD_i2r,zhang2024xwsb}, and ADD~\cite{ADD2023}. However, the use of a DR layer introduces challenges that limit the back-end’s ability to fully leverage the rich representations from speech foundation models. In this work, we aim to better unlock the potential of foundation models for speech anti-spoofing. These issues will be discussed in the next subsection.

\subsection{Limitation of Dimensionality Reduction Layer}
\label{sec_limitation_DR}

Existing speech foundation model-based anti-spoofing systems excel in extracting rich, high-dimensional feature representations, which capture intricate patterns in speech. However, this high dimensionality poses a significant challenge for downstream tasks. Models in these tasks typically require lower-dimensional features~\cite{AASIST, RawBMamba, li21o_interspeech}, creating a mismatch between the output features of the foundation models and the requirements of downstream processing.

A commonly used approach for dimensionality reduction is to employ a DR layer. However, this approach has several issues, including parameter overhead and potential information loss. As shown in Table~\ref{tab_percentage}, our analysis of back-end models further emphasizes the inefficiency of this approach.
We consider commonly used feature dimensions of $N=1024$ from large models~\cite{WavLM, wav2vec2}, and a reduced dimension of $D=128$, widely adopted in SOTA back-end models~\cite{hemlata_wav2vec2, SVDD_i2r, attentive_merge}. 

\begin{table}[t]
\centering
\caption{Contribution of the DR layer on the number of parameters and computational cost in back-end models. MMACs stands for million multiply-accumulate operations.}
\label{tab_percentage}
\begin{tabular}{l|ccc|crc}
\hline
\toprule
\multirow{2}{*}{Back-end Model} & \multicolumn{3}{c|}{Parameters} & \multicolumn{3}{c}{MMACs} \\ \cline{2-4} \cline{5-7}

 & DR &  Total & \% & DR &  Total & \% \\ \hline
\midrule
ResNet~\cite{resnet} & 131k & 611k & 21\% & 26.24 & 70.62 & 37\% \\
Res2Net~\cite{Res2Net} & 131k & 452k & 29\% & 26.24 & 64.93 & 40\% \\
ECAPA~\cite{ECAPA_TDNN} & 131k & 497k & 26\% & 26.24 & 80.21 & 33\% \\
AASIST~\cite{AASIST} & 131k & 447k & 29\% & 26.24 & 707.65 & 4\% \\
\bottomrule
\hline
\end{tabular}
\vspace{-0.1 in}
\end{table}

Across various back-end models, the DR layer, despite being just a single layer, consistently accounts for a substantial share of parameters and computational cost, underscoring its resource-intensive nature. For instance, the DR layer accounts for 21\% to 29\% of the parameters across ResNet, Res2Net, ECAPA, and AASIST. In terms of computational cost, the DR layer generally contributes at least one-third of the total MACs. AASIST is the only exception, where the DR layer accounts for just 4\% of the MACs, primarily because its overall MAC count is an order of magnitude higher than that of other models.

This table highlights that a single DR layer significantly inflates the back-end model’s size and resource demands. Furthermore, its direct projection design discards important high-dimensional features, limiting the overall potential of speech foundation models.

\section{Methodology}

\subsection{{Proposed Nested Res2Net (Nes2Net)}}

The design of Nes2Net is driven by two primary objectives: 1) effectively and efficiently utilizing the high-dimensional features from speech foundation models, and 2) enhancing multi-scale feature extraction to achieve robust generalization in speech anti-spoofing tasks. These objectives are realized through a novel nested architecture that simultaneously improves the efficiency, flexibility, and robustness of the model.

\textbf{Efficiency and Retention of Rich Feature Information:} The analysis in Section~\ref{sec_limitation_DR} reveals the limitations of employing the DR layer. {Building upon the observations,} Nes2Net entirely removes the DR layer, directly processing high-dimensional features to retain their intrinsic richness and minimize unnecessary computational costs.
By bypassing the DR layer, Nes2Net prevents the information bottleneck typically caused by early dimensionality reduction. This ensures the preservation of detailed representations essential for accurately distinguishing genuine speech from spoofed audio. 

\textbf{Enhanced Multi-Scale Feature Interaction and Expressiveness:}
While the Res2Net architecture effectively extracts multi-scale features through hierarchical splits, it exhibits significant limitations when processing high-dimensional features directly, especially with large split scales $s$. Specifically, Res2Net suffers from feature dilution~\cite{Res2Net}, redundant transformations~\cite{9383531}, and restricted interactions among channels. Excessive splitting fragments the features, weakening their expressiveness, and repetitive transformations increase computational redundancy, potentially causing overfitting. Moreover, closely related information can be distributed across non-adjacent subsets, limiting effective cross-channel interactions.

To overcome these limitations, as illustrated in Fig.~\ref{fig_Nes2Net},  we propose a novel \textbf{Nes}ted Res\textbf{2Net} (\textbf{Nes2Net}) architecture that introduces a hierarchical nesting structure. This additional degree of flexibility significantly enhances the model's representational capability. Each nested layer progressively refines features by building upon outputs from preceding layers and also incorporates efficient local cross-channel attention mechanisms~\cite{ECA_Net, MFA_TDNN}, strengthening interactions across channels. This holistic feature extraction approach enables Nes2Net to comprehensively capture intricate speech patterns. Moreover, the cumulative refinement effectively mitigates the issue of feature dilution, preserving rich and expressive multi-scale information. Benefiting from the structural advantages of the nesting strategy, the need for excessive fine-grained splits is reduced, effectively mitigating redundant transformations. This approach also minimizes unnecessary computations, resulting in a compact yet highly expressive model.

Critically, overfitting is a well-known challenge in speech anti-spoofing tasks, often leading to degraded performance in cross-domain scenarios. Previous studies~\cite{AASIST, li21o_interspeech}, particularly with compact models like AASIST and Res2Net (both with fewer than 500k parameters), have shown that smaller models can help reduce overfitting.  Our experiments with these models confirm that simply increasing their size does not always lead to better performance and can, in fact, make overfitting worse. As a result, improving feature quality through smarter model structure design becomes more important than just scaling up the model. The nested architecture of Nes2Net provides clear benefits as it maintains computational efficiency while reducing the risk of overfitting. 

The Nes2Net consists of an outer layer and several identical nested layers, described as follows:

\subsubsection{Outer Layer}
The outer layer of Nes2Net adopts a structure similar to that of Res2Net. The high-dimensional features produced by a speech foundation model are uniformly split into $s_1$ feature map subsets, denoted by $x_i$, where $i \in \{1, 2, \ldots, s_1\}$. Each feature subset $x_i$ has the same spatial size but contains only $\frac{1}{s_1}$ of the channels of the input feature map. With the exception of $x_1$, each $x_i$ is paired with a corresponding nested layer, denoted by $\mathbf{K}_i(\cdot)$. The output of $\mathbf{K}_i(\cdot)$, represented as $y_i$, is computed as follows: 

\begin{equation}
{y}_i =
\begin{cases} 
{x}_i & i = 1; \\ 
\mathbf{K}_i({x}_i) & i = 2; \\ 
\mathbf{K}_i({x}_i + {y}_{i-1}) & 2 < i \leq s_1.
\end{cases}
\label{eq_outer}
\end{equation}

where $x_i$ is first added to the output of $\mathbf{K}_{i-1}(\cdot)$, and the resulting feature map is then fed into $\mathbf{K}_i(\cdot)$ for further processing. All $y_i$ features are concatenated along the channel dimension. Due to the combinatorial explosion effect~\cite{Res2Net}, the output features encapsulate a fusion of receptive field characteristics across different scales and frame levels. These features are then pooled along the time axis to convert frame-level features into utterance-level representations, which are subsequently used to compute the final classification score.

It is worth noting that since the outer layer directly processes high-dimensional features from the speech foundation model, the original two convolutional layers (kernel size of $1$) used before splitting and after concatenation in Res2Net are removed to improve efficiency.

\subsubsection{Nested Layer}
 
The nested layer acts as the core module responsible for processing the outer layer’s intermediate features, denoted by $x'_i$, where $i \in \{2, \ldots, s_1\}$. Based on Eq.~\ref{eq_outer}, $x'_i$ is defined as:

\begin{equation}
{x'}_i =
\begin{cases} 
{x}_i & i = 2; \\ 
{x}_i + {y}_{i-1} & 2 < i \leq s_1.
\end{cases}
\end{equation}
Each nested layer $\mathbf{K}_i(\cdot)$ is designed to extract multi-scale representations from its input while maintaining computational efficiency.  As shown in Fig.~\ref{fig_Nes2Net}, the structure of  $\mathbf{K}_i(\cdot)$ follows a SE-Res2Net-like design, but its input is the feature subset $x'_i$ from the outer layer of Nes2Net. Specifically, each nested layer consists of the following components:  

\textbf{Convolutional Layers}: The input feature map is first processed by a convolutional layer with a kernel size of 1 to extract local features while preserving the spatial dimensions. 

\textbf{Multi-Scale Feature Extraction}: To enable multi-scale processing, the input feature map $x'_{i}$ is equally split into $s_2$ subsets along the channel dimension, denoted by $x'_{i,j}$, where $j \in \{1, 2, \ldots, s_2\}$. Each subset undergoes separate transformations through convolutional operations $\mathbf{M}_j$ with varying receptive fields, yielding $y_{i,j}$, formulated as:
\begin{equation}
{y}_{i,j} =
\begin{cases} 
{x'}_{i,j} & j = 1; \\ 
\mathbf{M}_j({x'}_{i,j} + {y}_{i,j-1}) & 1 < j \leq s_2.
\end{cases}
\end{equation}
These transformed subsets are then concatenated to form the output $y_i$ of the nested layer.

\textbf{SE Module}: To further enhance the feature representations, a Squeeze-and-Excitation (SE) module is integrated into each nested layer. The SE module adaptively recalibrates channel-wise features to emphasize informative features and suppress less relevant ones~\cite{SE}.

\textbf{Residual Connections}: To enhance gradient flow and stabilize training, a residual connection is applied by adding the input of $x'_i$ to its output $y_i$. This design preserves the original information while incorporating newly learned features.

{In summary, the nested layer is lightweight, highly efficient, and designed to improve robustness and generalization across diverse conditions.}

\subsection{Enhanced Nested Res2Net (Nes2Net-X)}
{Nes2Net efficiently addresses the high-dimensional feature issue. However, it relies on an additive combination method within the nested layer, which may limit the flexibility and effectiveness of feature extraction, as it implicitly assigns equal importance to all features.}
To further enhance the representational capacity of Nes2Net, we propose an improved variant named Nes2Net-X. {It replaces the original addition operation in the nested layer with a concatenation followed by a learnable weighted summation}. This design explicitly preserves feature subset individuality before fusion and employs learnable weights to adaptively combine these subsets.
The Nes2Net-X consists of the following components:

\textbf{Feature Splitting and Processing:} This component is the same as that in Nes2Net nested layer. The input feature $x'_{i}$ is equally split into $s_2$ subsets along the channel dimension, denoted by $x'_{i,j}$, where $j \in \{1, 2, \ldots, s_2\}$. Each subset $x'_{i,j}$ undergoes a convolutional operation to extract feature representations.

\textbf{Feature Concatenation}: The outputs of the convolutional layers are denoted as $z_{i,j}$. In Nes2Net-X, instead of summing the processed features as in the Nes2Net, each current subset $x'_{i,j}$ is concatenated with the previous output $z_{i,j-1}$ {along a newly introduced dimension} before being processed.

\textbf{Weighted Sum}: The additional dimension created during concatenation is merged back into the original feature space using a `weighted sum' operation. This operation enables the model to dynamically assign importance to each subset, enhancing feature representation.
For each subset, the `weighted sum' is applied to the output feature $z_{i,j}$ of the convolutional layer. Let $w_{i,j}$ denote the learnable weights assigned to each concatenated feature. The output $y_{i,j}$ of the `weighted sum' is computed as:
\begin{equation}
y_{i,j} = \sum_{k=1}^s w_{i,j,k} \cdot z_{i,j,k}
\end{equation}

where $s$ denotes the number of subsets, $w_{i,j,k}$ represents the weight for the $k$-th subset features $z_{i,j,k}$.

The weighted summation provides more flexible and effective feature integration, offering several advantages:

\begin{itemize}
    \item \textbf{Enhanced Feature Diversity}: By concatenating features across subsets, the network captures a richer set of features, encompassing various aspects of the input data. 
    \item \textbf{Learnable Feature Fusion}: The introduction of learnable weights $w$ enables the model to prioritize more informative features, effectively suppressing less relevant ones. This adaptive mechanism allows the network to focus on the most discriminative features for the task.
    \item  \textbf{Improved Gradient Flow}:  By combining concatenation with weighted summation, the model facilitates better gradient propagation during training. This helps address potential issues such as vanishing or exploding gradients, leading to more stable and efficient learning.
\end{itemize} 

These modifications enable Nes2Net-X to retain the strengths of Nes2Net while introducing greater flexibility in feature fusion, ultimately improving performance.

\section{Experimental Setups}
\subsection{Datasets}

\begin{table}[h]
\caption{An summary of the datasets used in our experiments.}
\scriptsize
\setlength{\tabcolsep}{0.85mm}{
\begin{tabular}{lcrrr}
\hline
\toprule
\label{table_datasets}
\multirow{2}{*}{Dataset} & \multirow{2}{*}{\begin{tabular}[c]{@{}c@{}}Spoofing \\ Type\end{tabular}} & \multicolumn{3}{c}{Number of Samples} \\ \cmidrule(r){3-5}
 &  & Train & Valid & Test \\
\hline
\midrule
CtrSVDD w/o ACESinger bona fide~\cite{CtrSVDD_dataset} & \multirow{2}{*}{Singing Voice} & \multirow{2}{*}{84,404} & \multirow{2}{*}{43,625} & 64,734 \\ 
CtrSVDD w/ ACESinger bona fide~\cite{CtrSVDD_dataset} &  &  &  & 67,579 \\ \hline
ASVspoof 2019~\cite{ASVspoof2019} & \multirow{5}{*}{Speech} & 25,380 & 24,844 & - \\
ASVspoof 2021 LA~\cite{ASVspoof2021} &  & - & - & 181,566 \\
ASVspoof 2021 DF~\cite{ASVspoof2021} &  & - & - & 611,829 \\
ASVspoof 5~\cite{ASVspoof5_data} &  & 182,357 & 140,950 & 680,774 \\
In-the-Wild~\cite{In_the_wild}      &  & - & - & 31,779  \\ 
\hline  

PartialSpoof~\cite{PartialSpoof_TASLP} &  Partial Spoof & 25,380 & 24,844 & 71,237\\
\bottomrule
\hline
\end{tabular}
}
\end{table}

We use five datasets across various scenarios, including singing voice deepfake, fully spoofed speech, adversarial attacks, and partially spoofed speech, to evaluate the performance of the proposed model.
Singing voice deepfake detection (SVDD) is a growing area of interest in the research community~\cite{SVDD_ICASSP24, CtrSVDD_dataset, 10446271}. {The CtrSVDD dataset~\cite{SVDD_ICASSP24, CtrSVDD_dataset} offers structured attack types and official evaluation protocols, making it suitable for systematic architecture exploration. As a newly collected resource, it captures recent spoofing techniques, providing a more challenging and relevant benchmark for modern anti-spoofing systems. We therefore adopt it as a representative example.}
Moreover, fully spoofed speech is the most studied category. In this work, we include two categories of datasets: (1) the ASVspoof series, which comprises ASVspoof 2019~\cite{ASVspoof2019}, ASVspoof 2021 Logical Access (LA), ASVspoof 2021 Deepfake (DF)~\cite{ASVspoof2021}, and ASVspoof 5~\cite{ASVspoof5_data}; and (2) the In-the-Wild dataset~\cite{In_the_wild}, which reflects real-world usage scenarios.
Partially spoofed speech alters only part of an utterance to convey deceptive meaning. This emerging challenge has attracted growing attention. We use the PartialSpoof~\cite{PartialSpoof_TASLP} dataset as a representative benchmark.
Table~\ref{table_datasets} summarizes the datasets used in this study. Models are trained on the training set and validated on the validation set to select the best checkpoint for testing.

For CtrSVDD~\cite{CtrSVDD_dataset}, we report results on two official test protocols, according to whether ACESinger bona fide samples are included. The `A14' {attack type of the CtrSVDD dataset} is excluded following the official guidelines~\cite{CtrSVDD_dataset}. 
ASVspoof 2019~\cite{ASVspoof2019} is used only for training and validation, while the In-the-Wild~\cite{In_the_wild}, ASVspoof 2021 LA and DF~\cite{ASVspoof2021} datasets are used only for testing. 
For the recently released ASVspoof 5 dataset~\cite{ASVspoof5_data}, we use its train, development, and evaluation partitions for model training, validation, and testing, respectively.
For PartialSpoof~\cite{PartialSpoof_TASLP}, we follow the standard partitioning into train, development, and evaluation sets.

\subsection{Training Strategies}
Each experiment is run three times using different random seeds. We report both the result from the best-performing run and the average performance across all runs.
The values of $s_1$ and $s_2$ are both set to 8 for Nes2Net and Nes2Net-X.
The baseline systems for each dataset are built using SOTA models, and our proposed model adopts similar training strategies. The details are as follows:

 \begin{figure}[h]
    \centerline{\includegraphics[scale=0.091]{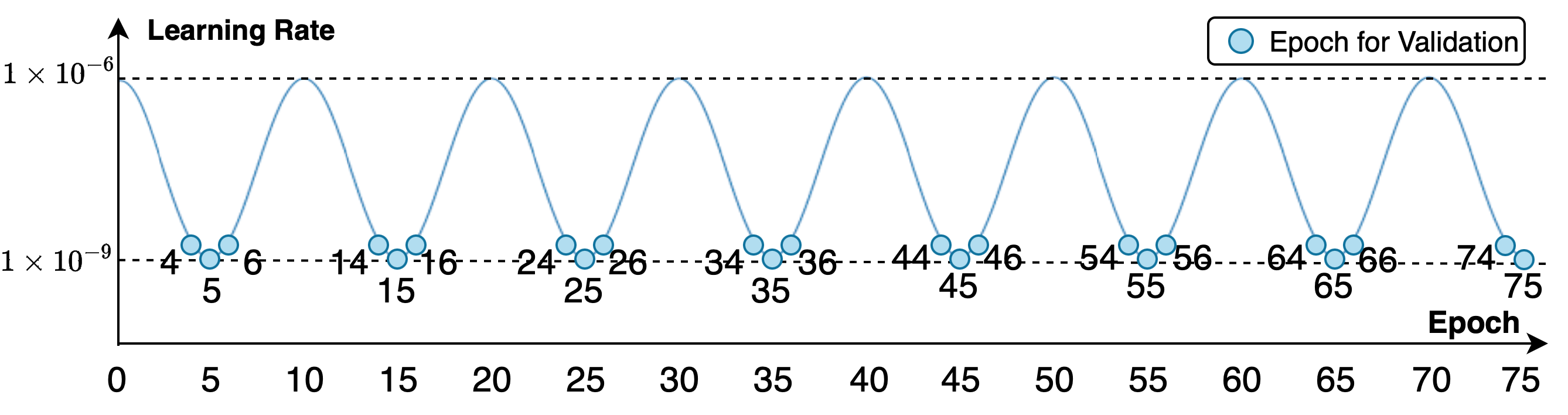}}
    \vspace{-0.1 in}
    \caption{The cyclic learning rate schedule using cosine annealing.}
    \label{fig_lr}
\end{figure}

\textbf{CtrSVDD:}
For the models trained on the CtrSVDD dataset~\cite{SVDD_ICASSP24, CtrSVDD_dataset}, we follow the baseline system from~\cite{SVDD_i2r}\footnote{\url{https://github.com/Anmol2059/SVDD2024}}. 
Following the setting in~\cite{SVDD_i2r}, we use a random seed of 42 to ensure reproducibility. Furthermore, due to the inherent stochasticity in deep learning, repeated runs are necessary to obtain reliable average results.
We use the AdamW optimizer with batch size 34, an initial learning rate of $1 \times 10^{-6}$, and weight decay of $1 \times 10^{-4}$. The learning rate is scheduled using cosine annealing with a cycle to a minimum of $1 \times 10^{-9}$.

\begin{table*}[t]
\centering
\caption{Performance in EER (\%) on the CtrSVDD evaluation set~\cite{CtrSVDD_dataset} with WavLM~\cite{WavLM} front-end. Results are shown as `best (mean)' over 3 runs. Params. and MMACs refer to number of parameters and million multiply-accumulate operations, respectively. w/o and w/ ACE B.F. refer to `without' and `with' ACESinger bona fide samples, respectively. Attack-specific EERs are computed under the `w/o ACE B.F' condition. Best results are in \textbf{bold}; second-best are \uline{underlined}.{`$\dag$' denotes implementation conducted by us.}}
\vspace{-0.05 in}
\setlength{\tabcolsep}{1.8mm}{
\begin{tabular}{lrrlllllll}
\hline
\toprule
\multirow{2}{*}{Back-end} & \multirow{2}{*}{Params.} & \multirow{2}{*}{MMACs} & \multicolumn{5}{c}{EER of Different Attack Types} & \multicolumn{2}{c}{Pooled EER} \\ \cmidrule(r){4-8} \cmidrule(r){9-10} 
 &  &  & A9 & A10 & A11 & A12 & A13 & \textbf{w/o ACE. B.F.} & w/ ACE. B.F. \\ 
 \hline
\midrule
{XWSB~\cite{zhang2024xwsb} ※} & - & - & - & - & - & - & - & -  & {2.32} \\ 

\hline

SLS~\cite{zhang2024xwsb}& - & - & - & - & - & - & - & -  & 2.59 \\ 

AASIST (C=32)~\cite{SVDD_i2r}& 447k & 707.65 & - & -& - & - & - &  - & 2.70 \\ \hline

AASIST Light (C=24)~$\dag$ & \textbf{159k} & 91.35 &  1.27 (1.37) &  0.87 (1.00) &  5.44 (5.86) &  4.84 (5.65)&  0.98 (1.05) &  3.95 (4.35) & 3.41 (3.77) \\
AASIST Standard(C=32)~$\dag$ & 447k & 707.65 &  \uline{1.18} (1.28) & 0.73 (0.86) & 3.63 (3.86) & 5.65 (5.77) & 0.88 (1.00) &  3.30 (3.36) & 2.79 (2.89) \\
AASIST Large(C=40)~$\dag$ & 662k & 1,091.28 & 1.32 (1.37) & 0.87 (0.97) & 3.70 (3.96) & 5.04 (5.63) & 0.96 (1.06) &  3.19 (3.36) & 2.71 (2.94) \\
AASIST XL(C=48)~$\dag$ & 835k & 1,555.56 & 1.23 (1.36)  &0.76 (0.92)  & 3.40 (4.64)  &  4.93 (5.55) &  0.89 (1.06) &  3.12 (3.62)   &2.76 (3.18)  \\
AASIST XXL(C=56)~$\dag$ & 1,087k & 2,104.57  &  \textbf{0.96 (1.20)} &  \uline{0.66} (0.84)&  3.86 (4.15) & 4.83  (5.43) & \textbf{0.75} \uline{(0.95)}  &  3.05 (3.43) & 2.65 (2.95) \\ 
\hline
ResNet~$\dag$ & 611k & 70.62 &  \uline{1.18 (1.21)} & 0.80 (0.93) & 3.97 (5.06) & \uline{4.60 (4.86)} & 0.96 (1.03) & 3.11 (3.61) & 2.74 (3.17) \\
Res2Net~$\dag$ & 452k & 64.93 &  1.26 (1.37)& 0.83 (0.86) & 3.59 (4.08) & \textbf{4.45 (4.80)} & 1.08 (1.09) &  3.02 (3.24) & 2.61 (2.78) \\
ECAPA-TDNN (C=128)~$\dag$ & 497k & 80.21 &  \uline{1.18} (1.39)& 0.67 (0.85) & 4.47 (5.84) & {4.63 (4.96)} &  0.87 (1.04) &  3.19 (3.74) & 2.79 (3.30) \\ 
\hline
\textbf{Proposed Nes2Net} & 511k & \textbf{58.11} &  1.23 (1.34)&  0.76 \uline{(0.81)} &  \uline{2.40 (2.43)} &  5.00 (5.24) &  0.96 (0.99) & \uline{2.53 (2.55)} & \uline{2.22 (2.27)} \\
\textbf{Proposed Nes2Net-X} & 511k & 91.35 &  {1.21} {(1.23)}&  \textbf{0.63 (0.76)}&  \textbf{2.09 (2.32)}&  4.99 (5.24)&  \uline{0.83} \textbf{(0.92)}& \textbf{2.48 (2.51)}& \textbf{2.20 (2.24)} \\

\bottomrule
\hline
\end{tabular}
}
\label{tab_SVDD}
\raggedright {※: XWSB is an ensemble-like model that combine two SSL front-ends~\cite{zhang2024xwsb}, while all other models in Table~\ref{tab_SVDD} are based on single SSL front-end.}
\vspace{-0.1 in}
\end{table*}

As shown in Fig.~\ref{fig_lr}, over 75 training epochs, we select checkpoints from the epoch with the minimum learning rate, as well as its preceding and following epochs, for validation. The best validation result is then used for testing. We use binary focal loss~\cite{lin2017focal}, a generalization of binary cross-entropy loss, with a focusing parameter ($\gamma$) of 2 and a positive class weight ($\alpha$) of 0.25. To standardize input length, each sample is randomly cropped or padded to 4 seconds during training. We adopt the Rawboost `parallel: (1)+(2)' data augmentation strategy~\cite{rawboost}, as explored in~\cite{SVDD_i2r}. WavLM is used as the front-end model for this dataset. The pre-trained and implementation of WavLM are obtained from S3PRL\footnote{\url{https://github.com/s3prl/s3prl}}.

\textbf{ASVspoof 2019 \& 2021:} For the models trained on the ASVspoof 2019~\cite{ASVspoof2019} dataset, we follow the baseline system proposed in~\cite{hemlata_wav2vec2}\footnote{\url{https://github.com/TakHemlata/SSL_Anti-spoofing}}. 
Audio data are cropped or concatenated to create segments of approximately 4 seconds in duration (64,600 samples) {for both training and testing}. We use the Adam optimizer~\cite{adam} with a weight decay of $1 \times 10^{-4}$. To reproduce the AASIST baseline~\cite{hemlata_wav2vec2}, we reduce the original batch size from 14 to 8 due to GPU memory constraints, and halve the learning rate from $1 \times 10^{-6}$ to $5 \times 10^{-7}$. For Nes2Net, benefiting from its lower GPU memory consumption, we use a batch size of 12 with a learning rate of $2.5 \times 10^{-7}$.
The loss function used is weighted Cross Entropy. Following~\cite{hemlata_wav2vec2}, we apply Rawboost augmentations~\cite{rawboost}, specifically `series: (1+2+3)' (Algo4) and `series: (1+2)' (Algo5), for AASIST baselines. For the proposed Nes2Net-X, only the former augmentation is applied. All models are trained for 100 epochs and the best checkpoint on the validation set is used for testing on the ASVspoof 2021~\cite{ASVspoof2021} and In-the-Wild~\cite{In_the_wild} datasets. 

\textbf{ASVspoof 5:} Both our AASIST baseline and the proposed Nes2Net-X models are trained using settings similar to those used for AASIST in the ASVspoof 2019 corpus. However, several differences apply. The final learning rate is set to $1 \times 10^{-7}$, we apply data augmentation using MUSAN \cite{musan} and RIR \cite{rir}, and training is stopped if there is no improvement on the development set for 5 consecutive epochs.

\textbf{PartialSpoof:} For models trained on the PartialSpoof~\cite{PartialSpoof_TASLP}, we follow the baseline systems described in~\cite{PartialSpoof_TASLP, 
how_do_partialspoof}\footnote{\url{https://github.com/nii-yamagishilab/PartialSpoof}}.
Specifically, we use wav2vec 2.0 as the front-end, the MSE for P2SGrad~\cite{wangxininterspeech} as the loss function, and Adam~\cite{adam} as the optimizer. Following~\cite{how_do_partialspoof}, the batch size is set to $2$, and a learning rate of $2.5 \times 10^{-6}$ is adopted for the baseline systems. For the proposed Nes2Net and Nes2Net-X, the learning rate is set to $1 \times 10^{-5}$.  The pooling layer used for the proposed Nes2Net and Nes2Net-X is the Attentive Statistics Pooling~\cite{Okabe2018}, and the reduction ratio of SE module is set to 8. Training is terminated if no improvement is observed on the development set for 20 consecutive epochs. The epoch yielding the best performance on the development set is used for testing.

\section{Results and Analysis}
All Equal Error Rate (EER) results in this work are reported as `best (mean)' over multiple runs. For cited results that (1) are based on a single run, (2) report only the best result, or (3) lack sufficient details, only a single value is presented.

\subsection{Studies on the CtrSVDD dataset}
\label{sec_ctrsvdd}

We conduct experiments on the CtrSVDD dataset~\cite{CtrSVDD_dataset}, following two testing protocols: one including ACESinger bona fide samples and the other excluding them~\cite{SVDD_report}. 
While results for both protocols are reported in Table~\ref{tab_SVDD},
our primary analysis focuses on the scenario `\textbf{without ACESinger bona fide (w/o ACE. B.F.)}', as recommended by the dataset creators. {Since AASIST (C=32) in our prior work~\cite{SVDD_i2r}, as well as SLS and XWSB~\cite{zhang2024xwsb}, were evaluated during the CtrSVDD Challenge 2024, portions of their test sets differ from the current official protocol. As a result, the EER by attack type is not directly comparable. To ensure a fair comparison, we re-implemented the AASIST (C=32) system under the official protocol and used it as our baseline, referred to as AASIST Standard (C=32) in Table~\ref{tab_SVDD}, achieving an EER of 2.79\%, which is close to the originally reported 2.70\%~\cite{SVDD_i2r}.}
Under the `w/o ACE B.F.' condition, the best run achieves an EER of 3.30\%, with an average of 3.36\% across three runs. Further experiments show that scaling up the AASIST model does not improve mean EER, possibly due to parameter redundancy.

\begin{table*}[t]
\centering
\caption{Performance in EER (\%) on the CtrSVDD evaluation set~\cite{CtrSVDD_dataset}, comparing the proposed Nes2Net with Res2Net and its various variants. The results are reported as the format of `best (mean)' across 3 runs, e.g., 3.02 (3.24) in the first row, or as the result of a single experiment, e.g., 3.21 in the second row. `b' and `s' represent the number of blocks and scale of Res2Net, respectively.}
\setlength{\tabcolsep}{1.7mm}{
\begin{tabular}{lccrrllc}
\hline
\toprule
\multirow{2}{*}{Back-end} & \multirow{2}{*}{\begin{tabular}[c]{@{}c@{}}Dimensionality \\ Reduction Layer\end{tabular}}   & \multirow{2}{*}{\begin{tabular}[c]{@{}c@{}}Reduced \\ Dimension $D$\end{tabular}}  
& \multirow{2}{*}{Params.} & \multirow{2}{*}{MMACs} & \multicolumn{2}{c}{Pooled EER} & \multirow{2}{*}{Remarks}  \\ \cmidrule(r){6-7}
 &  &  & &  & w/o ACE. B.F. & w/ ACE. B.F.\\ 
 \hline
\midrule
Res2Net (\textit{b=4, s=4}) & \checkmark & 128 & 452k & 64.93 &  3.02 (3.24) & 2.61 (2.78) & \multirow{4}{*}{\colorbox{teal}{\phantom{a}}  increase scale $s$} \\ 
Res2Net (\textit{b=4, s=16}) & \checkmark & 128 & 427k & 59.95 & 3.21  & 2.80 &  \\ 
Res2Net (\textit{b=4, s=64}) & \checkmark & 128 & 419k & 58.28 & 3.15  & 2.74 & \\
Res2Net (\textit{b=4, s=128}) & \checkmark & 128 & 417k & 57.98 &  3.26 & 2.88 & \\ \hline
Res2Net (\textit{b=4, s=4}) & \checkmark & 64 & 180k & 23.25 &  4.32  & 3.76 & \multirow{2}{*}{\colorbox{darkgrayblue}{\phantom{a}} change  $D$} \\ 
Res2Net (\textit{b=4, s=4}) & \checkmark & 256 & 1,273k & 202.91 &  3.83  & 3.38 & \\ \hline
Res2Net-woDR (\textit{b=1, s=4}) & \textbf{$\times$} & - & 861k & 119.15  & 4.15  & 3.62 & \multirow{7}{*}{\begin{tabular}[c]{@{}c@{}}\colorbox{grassgreen}{\phantom{a}} remove dimensionality \\ reduction layer \\ and increase scale $s$ \end{tabular}} \\
Res2Net-woDR (\textit{b=1, s=8}) & \textbf{$\times$} & -& 615k & 70.12 & 4.23 & 3.71 \\
Res2Net-woDR (\textit{b=1, s=16}) & \textbf{$\times$} & - & 456k & 38.24 &  3.82 & 3.35 \\
Res2Net-woDR (\textit{b=1, s=32}) & \textbf{$\times$} & - & 367k & 20.45 & 2.98 (3.45)  & 2.56 (3.02) \\
Res2Net-woDR (\textit{b=1, s=64}) & \textbf{$\times$} & - & 320k & 11.10 & 2.73 (2.97) & 2.42 (2.61) \\
Res2Net-woDR (\textit{b=1, s=128}) & \textbf{$\times$} & - & 296k & 6.31 &  3.29 & 2.88 \\ 
Res2Net-woDR (\textit{b=1, s=256}) & \textbf{$\times$} & - & 284k & 3.88 &  3.57 & 3.13 \\ \hline
Res2Net-woDR (\textit{b=2, s=64}) & \textbf{$\times$} & - & 637k & 21.78 & 3.20 & 2.82 & \multirow{2}{*}{\colorbox{lightred}{\phantom{a}} increase depth} \\
Res2Net-woDR (\textit{b=4, s=64}) & \textbf{$\times$} & - & 1,270k & 43.15 & 3.09 (3.18) & 2.73 (2.83) & \\
\hline
\textbf{Proposed Nes2Net} & \textbf{$\times$} & - & 511k & 58.11 & {2.53 (2.55)} & 2.22 {(2.27)} & \multirow{2}{*}{\colorbox{skyblue}{\phantom{a}} proposed nested design} \\
\textbf{Proposed Nes2Net-X} & \textbf{$\times$} & - & 511k & 91.35  & \textbf{2.48 (2.51)} & \textbf{2.20 (2.24)} \\
\bottomrule
\hline
\end{tabular}
}
\vspace{-0.1 in}
\label{tab_roadmap}
\end{table*}

We additionally evaluate several widely-used baseline systems, including ResNet~\cite{resnet}, Res2Net~\cite{Res2Net}, and ECAPA-TDNN~\cite{ECAPA_TDNN}. ECAPA-TDNN and ResNet achieve EERs of 3.74\% and 3.61\%, respectively, which are slightly worse than that of AASIST. In contrast, Res2Net benefits from the advantages of multi-scale feature extraction, delivering the best average performance among the baseline systems with an EER of 3.24\%. Our proposed Nes2Net outperforms all baseline systems, achieving {a mean} EER of 2.55\% with the lowest computational cost. Furthermore, the enhanced version, Nes2Net-X, further improves the performance to 2.51\% EER, marking the best single-model performance reported to date.
Compared to Res2Net, ResNet, ECAPA-TDNN, and SOTA AASIST ($C = 32$), Nes2Net-X achieves EER reductions of 23\%, 30\%, 33\%, and 25\%, respectively.

We also analyze performance across different synthetic attack types using the `w/o ACE B.F.' protocol. Except for the `A12' ~{attack type}~\cite{CtrSVDD_dataset}, our model consistently achieves either the best or second-best performance, demonstrating strong generalization and robustness.
Notably, the `A12' ~{attack type}, based on Singing Voice Synthesis (SVS), proves particularly challenging, showing higher EER across all models and highlighting a potential area for future improvement.

We observe that performance trends are consistent across both conditions, with and without ACESinger bona fide samples. Moreover, the EER is lower when ACESinger bona fide samples are included. This indicates that, even though ACESinger bona fide samples are considered out-of-domain, the trained models exhibit strong generalization capabilities and are able to classify these samples accurately.

\begin{figure}[t]
\includegraphics[width=\columnwidth]{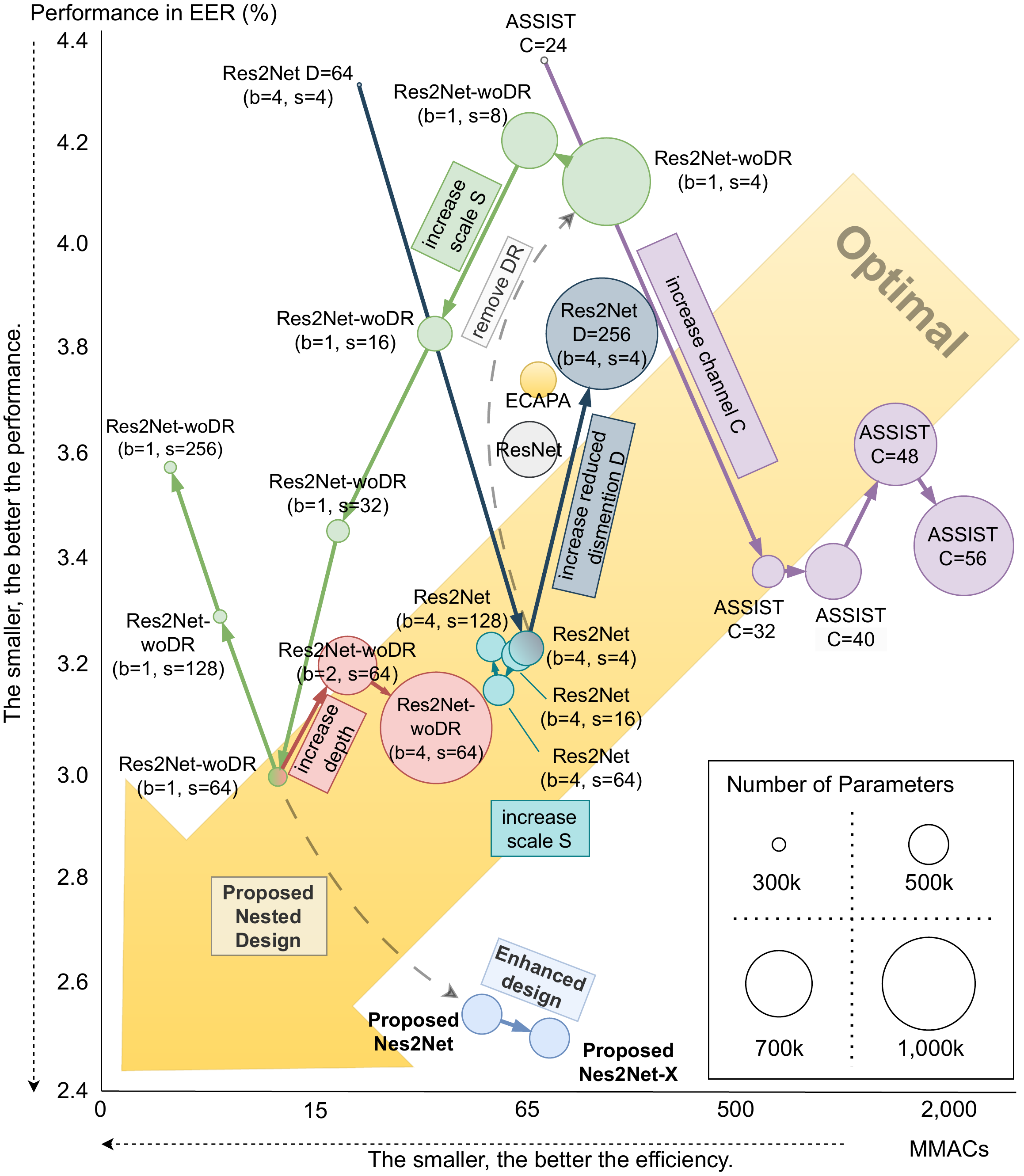}
\vspace{-0.25 in}
\caption{Visualization of Table~\ref{tab_SVDD} and~\ref{tab_roadmap}, highlighting our exploration of Res2Net and the roadmap of architectural changes leading to Nes2Net.}
\label{fig_roadmap}
\vspace{-0.15 in}
\end{figure}

\subsection{The Roadmap of the Nes2Net}
\label{sec_roadmap}
In this section, we introduce the roadmap from Res2Net to the proposed Nes2Net, with detailed results summarized in Table~\ref{tab_roadmap}. {All systems are implemented and evaluated under a unified framework for fair comparison.} To aid interpretation, we visualize the number of parameters, MACs, and EER. These are represented in Fig.~\ref{fig_roadmap} by circle size, the horizontal axis, and the vertical axis, respectively. In the following, we provide detailed analyses:

\textbf{Investigating Res2Net:}
Among the baselines in Table~\ref{tab_SVDD}, the Res2Net-based back-end outperforms ResNet, AASIST, and ECAPA-TDNN on the CtrSVDD dataset. Therefore, we select it as the reference baseline for further investigation.  First, we experiment with adjusting the scale $s$ of Res2Net. We observe that as $s$ increases, the number of split groups increases linearly; however, the performance shows no significant improvement (depicted as the teal blue line in Fig.~\ref{fig_roadmap}). This may be because adding too many split groups dilutes the feature representation, leading to redundancy.

Next, we explore varying the dimensionality of the output features from the DR layer (referred to as Reduced Dimension $D$, depicted as the steel gray line in Fig.~\ref{fig_roadmap}).
Reducing $D$ to 64 significantly lowers model size and MACs, compared to the default $D=128$, but leads to substantial performance degradation, increasing EER from 3.02\% to 4.32\%.
Conversely, increasing $D$ to 256 results in a much larger model size and MACs but still leads to worse performance than $D=128$. This may be because a larger $D$ introduces over-parameterization and noise. This may explain why $D=128$ is commonly adopted in SOTA models~\cite{hemlata_wav2vec2, SVDD_i2r}.

\textbf{Removal of DR Layer}:
Foundation models often incorporate a DR layer in their back-end architecture to compress high-dimensional features into lower-dimensional representations, facilitating downstream tasks. For instance, models like wav2vec 2.0-AASIST~\cite{hemlata_wav2vec2} utilize such a layer alongside task-specific classifiers (e.g., AASIST, ResNet). However, as discussed in Section~\ref{sec_limitation_DR}, this projection layer consumes a substantial portion of the back-end model's parameters and MACs while potentially causing information loss.

To explore whether bypassing this layer preserves more task-relevant information, we propose a new back-end model:  \textbf{ResNet} \textbf{w}ith\textbf{o}ut \textbf{D}imensionality \textbf{R}eduction (\textbf{ResNet-woDR}). By directly processing high-dimensional features, ResNet-woDR simplifies the architecture and focuses on the raw features extracted by the speech foundation model. The naming emphasizes the absence of a DR layer, differentiating it from traditional approaches.

We further evaluate the performance of ResNet-woDR with different scales $s$ (depicted as the green line in Fig.~\ref{fig_roadmap}). The best performance is observed with $s = 64$, achieving {a mean EER of 2.97\%}, which surpasses the best Res2Net baseline.
Increasing $s$ beyond this point leads to a decline in performance, likely due to the following factors:
\begin{itemize}
    \item \textbf{Feature Dilution.} A large $s$ excessively fragments feature representations, weakening their expressiveness and resulting in diluted, less informative features~\cite{Res2Net}. 
    \item \textbf{Redundant Transformations.} An overly large $s$ introduces unnecessary feature transformations, leading to overfitting and reduced generalization~\cite{9383531}.
    \item \textbf{Restricted Feature Interaction.} Since channels are unordered, distant groups may still contain correlated information. In this case, the additional convolutional layers introduced by splitting limit their interactions, weakening the model’s ability to capture complex patterns.
\end{itemize}
Based on the optimal $s$, we increase the number of blocks $b$ to deepen the model (depicted as the light pink line in Fig.~\ref{fig_roadmap}). However, no further performance improvement is observed. This could be attributed to the deeper architecture’s limited ability to effectively utilize the additional parameters, resulting in diminishing performance gains. It may also increase the risk of overfitting.

\textbf{The Novel Nested Design}:
Prior experiments demonstrate that removing the DR layer enhances the performance of Res2Net. We believe that directly extracting information from high-dimensional speech foundation model features avoids the information loss introduced by DR. Our experiments with variations in scale, depth, and dimensionality show that {a mean EER of 2.97\%} marks a performance bottleneck for this design. 

Compared to ResNet-woDR, the proposed Nes2Net adopts a novel nested design that enhances flexibility and significantly boosts the model's representational capacity. Processing larger feature subsets in the outer layer facilitates better interactions across channels within each nested layer. 
Furthermore, the integrated local cross-channel attention mechanism enhances feature selection while mitigating redundancy, addressing limitations in prior designs.
This architectural refinement overcomes the performance limitations observed in the original Res2Net design. As a result, Nes2Net and its enhanced variant Nes2Net-X surpass the earlier performance bottleneck, achieving {mean} EERs of 2.55\% and 2.51\%, respectively.

\subsection{Studies on the ASVspoof 2021 dataset}

\begin{table}[h]
\scriptsize
\centering
\vspace{-0.15 in}
\caption{Performance in EER (\%) on the ASVspoof 2021 LA and DF. The results are reported in the format of `best (mean)'. CKPT Avg. refers to the number of checkpoints averaged. {`$\dag$' denotes re-implementation conducted by us}. `Algo4' and `Algo5' represent Rawboost series augmentations: `(1+2+3)' and `(1+2)'~\cite{rawboost}, respectively. Parameters that are underlined are calculated by us. `-' represents unknown.
`N/A' indicates that the system does not use the average checkpoints method.} 
\vspace{-0.05 in}
\setlength{\tabcolsep}{0.5mm}{

\begin{tabular}{lllrcll}
\hline
\toprule
\multirow{2}{*}{Remark} & \multirow{2}{*}{Front-end}   & \multirow{2}{*}{\begin{tabular}[c]{@{}c@{}}Back-end \\ Model \end{tabular}}  & \multirow{2}{*}{\begin{tabular}[c]{@{}c@{}}Back-end \\ Parameters \end{tabular}}    & \multirow{2}{*}{\begin{tabular}[c]{@{}c@{}}CKPT \\ Avg.\end{tabular}} & \multicolumn{2}{c}{ASVspoof 2021}   \\ \cmidrule(r){6-7} 
 &  & &  & & LA & \textbf{DF} \\
\hline
\midrule
2022 & wav2vec 2.0 & FIR-NB~\cite{9747768}            &  -{\ \ }  &  -  &  3.54 & 6.18  \\
2022 & wav2vec 2.0 & FIR-WB~\cite{9747768}           &   -{\ \ }  & -   &  7.08 & 4.98  \\
2022 & wav2vec 2.0 & LGF~\cite{wang2021investigating} & -{\ \ }  & -  &  9.66 & 4.75  \\

2023 & wav2vec 2.0 & Conformer (fix)~\cite{rosello23_interspeech} & {2,506k}\tablefootnote{\url{https://github.com/ErosRos/conformer-based-classifier-for-anti-spoofing}}{\ } & {5}  &  1.38 & 2.27  \\
2023 & wav2vec 2.0 & Conformer (var)~\cite{rosello23_interspeech} & {2,506k}{\ \ } & {5}  &  0.87 & 7.36  \\
2024 & wav2vec 2.0 & Ensembling~\cite{rosello24_interspeech} $\ddagger$ & -{\ \ }  & -  &  2.32 (4.48) & 5.60 (8.74) \\
2024 & WavLM & ASP+MLP~\cite{tran24_interspeech}& \uline{1,051k}{\ \ } & - & 3.31& 4.47 \\   
2024 & wav2vec 2.0 & SLIM~\cite{SLIM} & -{\ \ } & - & - & {{\ \ - \ \ (4.4)}} \\
2024 & WavLM & AttM-LSTM~\cite{attentive_merge}& 936k\tablefootnote{\url{https://github.com/pandarialTJU/AttM_INTERSPEECH24}}{\ } & N/A & 3.50 & 3.19 \\
2024 & wav2vec 2.0 & FTDKD~\cite{FTDKD}      &  -{\ \ } &  -   & 2.96 & 2.82   \\
2024 & wav2vec 2.0 & AASIST2~\cite{10448049}         &  -{\ \ }  &   -   & 1.61 &2.77 \\
2024 & wav2vec 2.0 & MFA~\cite{10447923}          & -{\ \ } & - &  5.08 & 2.56  \\
2024 & wav2vec 2.0 & MoE~\cite{MOE} & -{\ \ } &  - &  2.96 & 2.54  \\
2024 & wav2vec 2.0 & OCKD~\cite{10446270} & -{\ \ } &  - &  0.90 & 2.27  \\

2024 & wav2vec 2.0 & TCM~\cite{TCM}      &  2,383k\tablefootnote{\url{https://github.com/ductuantruong/tcm_add}}{\ }  &  5   & 1.03  & 2.06   \\
2024 & wav2vec 2.0 & SLS~\cite{SLS} & {23,399k}\tablefootnote{\url{https://github.com/QiShanZhang/SLSforASVspoof-2021-DF}}{\ } 
& - & 2.87 (3.88) & 1.92 (2.09)  \\
2025 & wav2vec 2.0 & LSR+LSA~\cite{LSR+LSA} & -{\ \ } & - & 1.19 & 2.43 \\
2025 & wav2vec 2.0 & LSR+LSA~\cite{LSR+LSA} ※ & -{\ \ } & - & 1.05 & 1.86 \\

2025 & wav2vec 2.0 & WaveSpec~\cite{WaveSpec} & -{\ \ } & - & - & 1.90 \\
2025 & wav2vec 2.0 & Mamba~\cite{Mamba}     &   {1,937k}\tablefootnote{\url{https://github.com/swagshaw/XLSR-Mamba}}{\ }   &   5    & 0.93  & 1.88  \\
2025 & wav2vec 2.0 & {SSL-EOW-S.}~\cite{confen} $\ddagger$ & -{\ \ } & - & - & {1.75 (2.91)}\\
2025 & wav2vec 2.0 & {Cal. Ensemble}~\cite{confen} $\ddagger$ & -{\ \ } & - & - & {\ \ - \ \ (2.03)}\\
\hline
2022 & wav2vec 2.0 & AASIST~\cite{hemlata_wav2vec2}  &  447k\tablefootnote{\url{https://github.com/TakHemlata/SSL_Anti-spoofing}}
& N/A &  \textbf{0.82 (1.00)}  & 2.85 (3.69) \\ 
$\dag$  &  wav2vec 2.0 & AASIST (algo4) & 447k{\ \ } & N/A & 1.13 (1.36) & 3.37 (4.09) \\
$\dag$  &  wav2vec 2.0 & AASIST (algo5) & 447k{\ \ } & N/A & 0.93 (1.40) & 3.56 (5.07)   \\
\hline

\multirow{3}{*}{Ours} & wav2vec 2.0 & \textbf{{Nes2Net}} & {511k}{\ \ } & {N/A} & {1.61 (1.90)} & {1.89 (2.12)} \\ 
 & wav2vec 2.0 & \textbf{Nes2Net-X} & 511k{\ \ } & N/A & 1.73 (1.95) & 1.65 (1.91) \\ 
& wav2vec 2.0 & \textbf{Nes2Net-X} & 511k{\ \ } & 3 & 1.66 (1.87) & 1.54 (1.98) \\ 
& wav2vec 2.0 & \textbf{Nes2Net-X} & 511k{\ \ } & 5 & 1.88 (2.00) & \textbf{1.49 (1.78)} \\ 

\bottomrule
\hline
\end{tabular}
}
\raggedright ※: with extra data augmentation~\cite{LSR+LSA}

\raggedright $\ddagger$: ensemble of multiple models
\label{tab_asvspoof}
\vspace{-0.05 in}
\end{table}

\begin{table*}[t]
\scriptsize
\centering
\caption{Performance in EER (\%) for different types of vocoders and compression conditions on the ASVspoof 2021 DF test set. The five EER values for each sub-item, from left to right, correspond to Nes2Net-X, Mamba~\cite{Mamba}, SLS~\cite{SLS}, TCM~\cite{TCM}, and AASIST~\cite{hemlata_wav2vec2}.The best performance is reported in bold fonts, and the second-best is underlined.}
\vspace{-0.05 in}
\label{tab_21DF}
\renewcommand\arraystretch{1.1}
\setlength{\tabcolsep}{1.45mm}{
\begin{tabular}{lcccccc}
\hline
\toprule
             & Traditional Vocoder                       & Wav Concatenation                                 & Neural Autoreg.                            & Neural Non-autoreg.                       & Unknown                                           & Pooled EER                                 \\
             \hline
\midrule
C1 –         & \textbf{0.36}/\uline{0.78}/1.21/0.95/1.22 & \textbf{0.76}/\textbf{0.76}/0.80/\textbf{0.76}/2.28        & \textbf{2.70}/3.88/\uline{3.12}/3.89/3.45  & \textbf{0.52}/0.87/\uline{0.68}/0.95/1.56 & 1.64/\uline{1.63}/\textbf{1.23}/1.73/1.99         & \textbf{1.47}/1.89/\uline{1.72}/2.23/2.34  \\
C2 Low mp3   & \uline{1.48}/\textbf{0.94}/1.94/1.67/2.72 & 2.96/\uline{2.20}/\textbf{2.16}/2.56/5.84         & \uline{2.89}/3.23/\textbf{2.71}/3.59/5.96 & 1.23/\uline{0.86}/\textbf{0.78}/1.32/3.33 & 2.54/\uline{1.69}/\textbf{1.65}/1.93/4.30         & \textbf{1.75}/\uline{1.84}/2.02/2.11/4.30  \\
C3 High mp3  & \textbf{0.44}/\uline{0.88}/1.39/0.96/1.83 & \textbf{1.13}/1.49/\uline{1.17}/1.45/3.35         & \textbf{2.47}/3.35/\uline{2.91}/3.70/3.79  & \textbf{0.44}/0.87/\uline{0.69}/0.88/2.02 & 2.29/1.85/\textbf{1.34}/\uline{1.67}/2.65         & \textbf{1.32}/1.85/\uline{1.59}/1.95/2.64  \\
C4 Low m4a   & \textbf{0.44}/\uline{0.95}/1.48/1.22/1.57 & \uline{1.15}/\textbf{0.85}/1.24/1.67/2.09         & \textbf{2.79}/3.39/\textbf{2.79}/3.40/3.75 & \textbf{0.54}/0.96/\uline{0.70}/1.22/1.65 & 1.32/\uline{1.22}/\textbf{1.14}/1.41/2.10         & \textbf{1.40}/1.92/\uline{1.74}/2.01/2.37  \\
C5 High m4a  & \textbf{0.45}/\uline{0.80}/1.34/0.98/1.16 & \textbf{0.62}/0.76/\uline{0.71}/0.76/2.10         & \textbf{2.77}/3.48/\uline{2.96}/3.73/3.39  & \textbf{0.56}/0.90/\uline{0.64}/1.07/1.34 & 1.88/1.70/\textbf{1.34}/\uline{1.43}/1.87         & \textbf{1.59}/2.05/\uline{1.79}/1.96/2.14  \\
C6 Low ogg   & \textbf{0.69}/\uline{1.13}/2.14/1.44/2.35 & \textbf{0.80}/0.97/\uline{0.91}/\uline{0.91}/2.23 & \textbf{1.92}/2.80/\uline{2.44}/2.79/3.67  & \textbf{0.48}/0.78/\uline{0.61}/0.84/1.62 & 1.05/1.14/\textbf{1.00}/\uline{1.01}/2.23         & \textbf{1.09}/\uline{1.61}/1.88/1.87/2.58  \\
C7 High ogg  & \textbf{0.70}/\uline{1.13}/1.52/1.35/1.57 & \textbf{0.62}/0.80/\uline{0.71}/0.80/1.50         & \textbf{2.05}/2.84/\uline{2.26}/2.66/2.92  & \textbf{0.43}/0.65/\uline{0.52}/0.74/1.00 & 1.34/1.05/\textbf{0.96}/\textbf{0.96}/1.27        & \textbf{1.35}/1.61/\uline{1.57}/1.74/1.92  \\
C8 mp3→m4a  & \textbf{0.95}/\uline{1.26}/2.28/1.74/3.01 & 1.52/\textbf{0.97}/\uline{1.08}/\uline{1.08}/2.96 & \textbf{2.22}/3.01/\uline{2.31}/2.96/4.49  & \uline{0.61}/\textbf{0.57}/0.65/0.95/2.05 & 1.61/\uline{1.18}/\textbf{1.09}/\uline{1.18}/2.66 & \textbf{1.48}/\uline{1.65}/1.92/1.97/3.31  \\
C9 ogg→m4a & \textbf{0.70}/\uline{1.26}/2.15/1.49/2.28 & \textbf{0.88}/0.97/0.99/\textbf{0.88}/2.52         & \textbf{1.92/}3.01/\uline{2.57}/2.88/3.76  & \textbf{0.52}/0.70/\uline{0.65}/0.78/1.57 & \textbf{0.96}/1.09/1.09/\uline{1.05}/2.14         & \textbf{1.13}/\uline{1.79}/2.04/1.88/2.75  \\
Pooled EER   & \textbf{0.72}/\uline{1.14}/1.88/1.40/2.15 & 1.10/\textbf{1.05}/\uline{1.07}/1.14/2.85         & \textbf{2.70}/3.32/\uline{2.86}/3.40/4.05  & \textbf{0.63}/0.80/\uline{0.69}/0.94/1.84 & 1.86/1.43/\textbf{1.23}/\uline{1.38}/2.45         & \textbf{1.49}/\uline{1.88}/1.92/2.06/2.85 \\
\bottomrule
\hline
\end{tabular}
}
\vspace{-0.05 in}
\end{table*}
\begin{figure*}[h]
\centering
\includegraphics[width=0.93\textwidth]{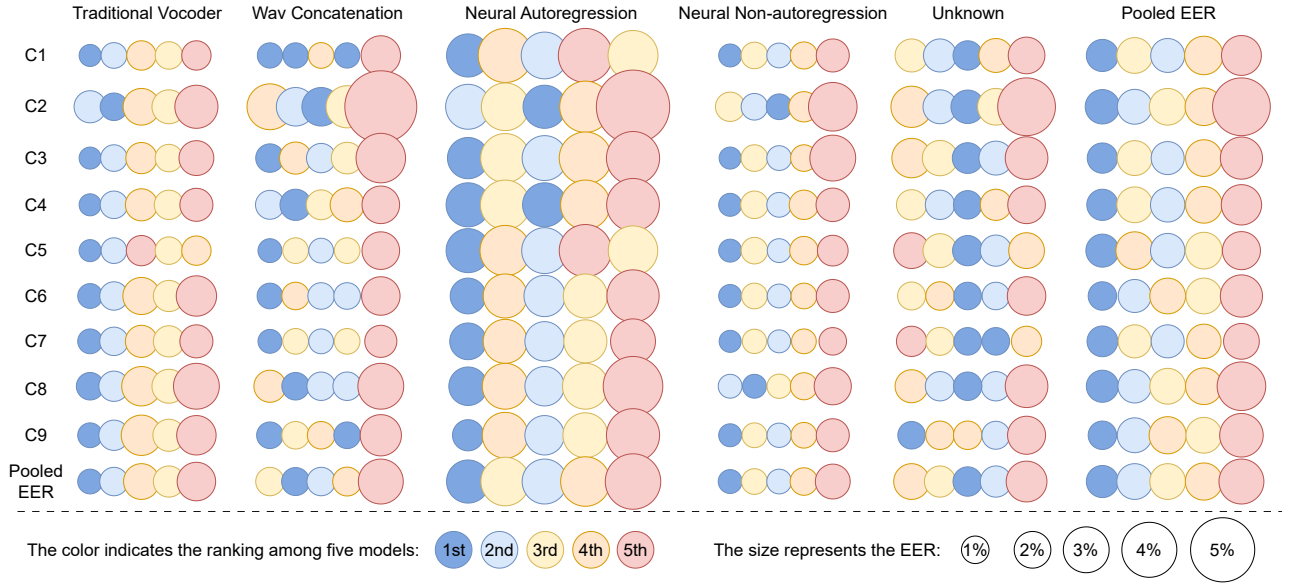}
\vspace{-0.1 in}
\caption{Visualization of the EER (\%) across various vocoders and compression conditions on the ASVspoof 2021 DF test set. Each EER value is shown as a colored circle, where the size indicates the EER value, and the color represents the performance ranking among the five models: blue (best) to light red (worst). The five EER values for each sub-item, from left to right, correspond to the proposed Nes2Net-X, Mamba~\cite{Mamba}, SLS~\cite{SLS}, TCM~\cite{TCM}, and AASIST~\cite{hemlata_wav2vec2}.}
\label{fig_21DF}
\vspace{-0.1 in}
\end{figure*}

The ASVspoof series datasets are widely used as benchmarks for advancing research in detecting spoofed speech~\cite{ASVspoof2019, ASVspoof2021}. Following the standard protocol, we train models on ASVspoof 2019~\cite{ASVspoof2019} and evaluate them on ASVspoof 2021 Logical Access (LA) and Deepfake (DF) tasks~\cite{ASVspoof2021}.
The LA task focuses on detecting synthetic and voice-converted speech transmitted over telephony systems, introducing challenges related to channel effects and transmission variability. In contrast, the DF task targets detecting manipulated, compressed speech data commonly found on online platforms. This reflects real-world scenarios where deepfake audio circulates, making the DF task a valuable benchmark for evaluating deepfake detection systems.

The results in Table~\ref{tab_asvspoof} show that for the LA track, our Nes2Net achieves {a mean EER of 1.90\%}, comparable to SOTA systems. For the DF track, which more closely reflects real-world scenarios as discussed earlier, the baseline system AASIST~\cite{hemlata_wav2vec2} achieves {its best EER of 2.85\% and a mean EER of 3.69\%}, remaining competitive with current SOTA systems. The SLS~\cite{SLS} and TCM~\cite{TCM} models achieve EERs close to 2\%, demonstrating strong performance at the SOTA level. The Mamba-based~\cite{Mamba} model further improves results, reducing the EER to 1.88\%. Notably, {our proposed Nes2Net attains its best EER of 1.89\% and a mean EER of 2.12\% EER}, comparable to the performance of current SOTA systems. The enhanced variant, Nes2Net-X achieves the best performance among all compared systems, with {its best EER of 1.65\% and a mean EER of 1.91\%}. 

Inspired by prior works~\cite{TCM, Mamba}, we average the weights of several top-performing checkpoints on the validation set to obtain an improved model. This approach further improves the performance of the DF task to {a best EER of 1.49\% and a mean EER of 1.78\%}, which, to the best of our knowledge, is the best performance reported to date. Furthermore, compared to Mamba~\cite{Mamba}, our model achieves this performance with approximately 74\% fewer parameters, demonstrating superior efficiency.

The analysis above summarizes overall performance on the DF test set. The DF dataset also provides detailed labels for vocoder types and compression conditions, enabling more fine-grained analysis.
To further evaluate performance, we compare the SOTA models Mamba, SLS, TCM, and AASIST with our proposed Nes2Net-X across these sub-tracks. The results are presented in Table~\ref{tab_21DF}.
To improve readability and make the extensive numerical data easier to interpret, we also visualize the table’s results in Fig.~\ref{fig_21DF}.

For traditional vocoders, all models perform well, with most EERs below 2\%. Notably, our proposed Nes2Net-X achieves exceptional results, consistently yielding EERs under 1\% across all conditions except C2. This demonstrates the strong stability of Nes2Net-X when handling unseen and relatively simple scenarios. In contrast, for neural autoregressive vocoders, all models experience a noticeable drop in performance, with EER reaching up to 5.96\%. This indicates the greater challenge posed by the sequential and dynamic nature of autoregressive vocoders, which introduce higher variability in synthesis. Nevertheless, Nes2Net-X maintains a clear advantage over the competing models, demonstrating its robustness in handling these complex synthesis conditions.

From the perspective of compression conditions, the differences in model performance are less pronounced compared to the variations observed across vocoder types. Nes2Net-X consistently achieves the lowest EERs across all compression conditions, regardless of the level of distortion introduced by compression. This consistency highlights the model’s strong generalization ability across different levels of compressions.

Overall, these findings demonstrate that Nes2Net-X is not only highly effective across diverse vocoder types, but also maintains superior performance under varying compression conditions. This robustness underscores the model’s capability to handle both compression diversity and complex synthesis challenges, making it a reliable solution for deepfake audio detection across a wide range of scenarios.

\subsection{The results on the In-the-Wild dataset}

\begin{table}[h]
\vspace{-0.15 in}
\caption{Performance in EER (\%) on the In-the-Wild~\cite{In_the_wild} dataset. Our result is reported as the format of `best (mean)' across 3 runs. }
\vspace{-0.05 in}
\label{tab_ITW}
\centering
\begin{tabular}{ccll}
\hline
\toprule
Front-end  & Year & \multicolumn{1}{c}{{Back-end}}  & \multicolumn{1}{c}{EER} \\ 
\hline
\midrule
\multirow{13}{*}{\rotatebox{90}{wav2vec 2.0}} & 2022 & Rawnet\&ASSIST {(reported by~\cite{SLS})} & 10.46 \\
 & 2024 & SLIM~\cite{SLIM} & {{\ \ - \ \ (12.5)}} \\
 &2024& MoE~\cite{MOE} & 9.17 \\
 & 2024& Conformer~\cite{rosello23_interspeech} & 8.42 \\
 & 2024& TCM~\cite{TCM}  & 7.79 \\
 & 2024  & OCKD~\cite{10446270} & 7.68  \\
 & 2024& SLS~\cite{SLS}  & 7.46 {(8.87)} \\
  & 2024& Pascu et al.~\cite{pascu24_interspeech}  & {{\ \ - \ \ (7.2)}}\\
 
 &2025 & Mamba~\cite{Mamba} & 6.71 \\
  &2025 & WaveSpec~\cite{WaveSpec} & 6.58 \\
 
 & 2025 & LSR+LSA~\cite{LSR+LSA} & 5.92  \\  
  & 2025 & LSR+LSA~\cite{LSR+LSA} ※ & 5.54 \\ 
 \cline{2-4}
  & - & \textbf{{Proposed Nes2Net}} & {{5.80 (7.06)}} \\
 & - & \textbf{Proposed Nes2Net-X} & \textbf{5.52 (6.60)} \\
\bottomrule
\hline
\end{tabular}

\raggedright \ \ \ \ \ \ \ \ \ ※: with extra data augmentation~\cite{LSR+LSA}
\vspace{-0.1 in}
\end{table}

The In-the-Wild dataset~\cite{In_the_wild} is a collection of deepfake videos sourced from the internet. Unlike controlled datasets, it captures the diverse and unpredictable nature of real-world scenarios. This diversity is essential for developing and evaluating deepfake detection models, as it challenges them to generalize effectively across a wide range of conditions.

In addition, unlike many other datasets that rely on self-generated fake audio, this dataset is collected from publicly available video and audio files explicitly labeled as audio deepfakes~\cite{In_the_wild}. To account for the potential presence of partial spoofing, we evaluate our proposed Nes2Net and Nes2Net-X using the entire duration of each test sample instead of restricting it to the first 4 seconds, as the latter approach risks missing partially spoofed segments.

The testing results, alongside SOTA models, are reported in Table~\ref{tab_ITW}. We find that the overall performance trends are consistent with those seen on the ASVspoof 2021 DF dataset. However, EERs on the In-the-Wild dataset are generally higher than those on the DF dataset, reflecting greater complexity and variability in real-world scenarios. Notably, the proposed Nes2Net-X outperforms all SOTA models, achieving the lowest EER of 5.52\% {and a mean EER of 6.60\%} on this challenging dataset.

\subsection{The results on the ASVspoof 5 dataset} 

\begin{table}[h]
\vspace{-0.1 in}
\centering
\caption{A comparison between the proposed Nes2Net and the AASIST baseline system on the ASVspoof 5 dataset~\cite{ASVspoof5_data}. `Params.' and `MMACs' refer to the number of parameters and the number of million multiply-accumulate operations, respectively. `Avg.' indicates the average relative performance improvement across all three evaluation metrics.}
\vspace{-0.05 in}
\label{tab_asvspoof5}
\renewcommand\arraystretch{1.1}
\setlength{\tabcolsep}{1.0mm}{
\begin{tabular}{ccccccc}
\hline
\toprule
 \multicolumn{3}{c}{Back-end} & \multicolumn{4}{c}{Performance} \\ \cmidrule(r){1-3} \cmidrule(r){4-7}
 {Model} & Params.↓ & MMACs↓ & CLLR↓ &  minDCF↓ & EER↓ & \textbf{Avg.} \\ 
 \hline
\midrule
 \multirow{1}{*}{AASIST} & \textbf{447k} & 707.65 & 0.9587  &  0.1645 & 6.08 & {\scriptsize Benchmark} \\ 
\textbf{{Nes2Net}} & {511k} & \textbf{{58.11}} & {0.7912} & {0.1568} & {6.13} & \textbf{{7.1\%}}\\
\textbf{Nes2Net-X} & 511k & {91.35} & \textbf{0.7344} & \textbf{0.1535} & \textbf{5.92} & \textbf{10.9\%}\\

\bottomrule
\hline
\end{tabular}
}
\vspace{-0.1 in}
\end{table}

The ASVspoof 5 dataset represents the most recent edition in the ASVspoof series. Unlike earlier versions, it introduces adversarial attacks and is crowdsourced under various acoustic conditions~\cite{ASVspoof5_data}. As it is newly released, there are currently no existing systems available for a fair comparison. Therefore, {we re-implement the AASIST system as a baseline and compare it with our proposed Nes2Net and Nes2Net-X model}. Following the ASVspoof 5 challenge guidelines \cite{ASVspoof5_data}, we use WavLM~\cite{WavLM} as the front-end. Based the evaluation protocol in~\cite{ASVspoof5_report}, we assess performance using three metrics: Cost of Log-Likelihood Ratio (CLLR), minimum Detection Cost Function (minDCF), and EER, and present the results in Table \ref{tab_asvspoof5}. We observe that the Nes2Net and Nes2Net-X back-end models result in only a slight increase in the number of parameters compared to AASIST, while significantly reducing MMACs. Moreover, across all three evaluation metrics, {{the Nes2Net and Nes2Net-X back-ends improve performance by 7.1\% and 10.9\%, receptively}.}

\subsection{The results on the PartialSpoof dataset}

\begin{table}[h]
\vspace{-0.1 in}
\caption{Performance in EER(\%) on the PartialSpoof~\cite{PartialSpoof_TASLP} dataset. The results are reported as the format of `best (mean)' across 3 runs. \dag \ indicates results obtained from our implementation.}
\vspace{-0.05 in}
\label{tab_partialspoof}
\centering
\begin{tabular}{cllll}
\hline
\toprule
\multicolumn{1}{c}{\multirow{2}{*}{Front-end}}  & \multicolumn{1}{c}{\multirow{2}{*}{Year}}  & \multicolumn{1}{c}{\multirow{2}{*}{Back-end}} & \multicolumn{2}{c}{PartialSpoof~\cite{PartialSpoof_TASLP}} \\ \cmidrule(r){4-5}
\multicolumn{1}{c}{}& & & Dev & Eval \\
\hline
\midrule
\multirow{7}{*}{\rotatebox{90}{wav2vec 2.0}} & 2024 &gMLP~\cite{PartialSpoof_TASLP} & 0.35 & {0.64} \\
 & - &gMLP\dag & 0.39 (0.43) & 0.72 (0.80) \\
&2024 &1D Res2Net~\cite{how_do_partialspoof} & 0.35 & 0.73 \\
& - &1D Res2Net\dag & 0.35 (0.38)  & 0.73 (0.79) \\
& - &SE ResNet\dag & 0.31 (0.50) & 0.77 (0.78) \\
\cline{2-5}
& - &\textbf{Nes2Net} & 0.24 {(0.36)} & \textbf{0.53} (0.68) \\
& -&\textbf{Nes2Net-X} & \textbf{0.20 (0.33) } & 0.57 \textbf{(0.64)} \\
\bottomrule
\hline
\vspace{-0.05 in}
\end{tabular}
\end{table}

Partially manipulating a sentence can significantly alter its intended meaning~\cite{how_do_partialspoof}. When such manipulations occur in small regions, existing models trained on fully spoofed speech and relying on pooling functions struggle to detect these subtle changes. Consequently, there is growing interest in the detection of partially spoofed speech~\cite{how_do_partialspoof, PartialSpoof_TASLP, LlamaPartialSpoof}.

To evaluate the performance of our proposed model across different spoofing tasks, we conduct experiments on the PartialSpoof dataset~\cite{PartialSpoof_TASLP}. The results are presented in Table~\ref{tab_partialspoof}.
First, we reproduce the performance of two SOTA models, achieving results comparable to those reported in their original papers~\cite{how_do_partialspoof, PartialSpoof_TASLP}. Additionally, we evaluate SE ResNet, which demonstrated performance similar to the other baselines. In contrast, our proposed Nes2Net and Nes2Net-X outperform all three baselines.

\begin{table*}[t]
\caption{The performance in EER (\%) on the ASVspoof 2021 LA, DF~\cite{ASVspoof2021}, and In-the-Wild~\cite{In_the_wild} datasets. The results are reported as the format of `best (mean)' across 3 runs. `w/ Aug.' and `w/o Aug.' indicate whether evaluation with augmentations on the validation set is used to select the best checkpoint for testing. CKPT Avg. refers to the number of checkpoints averaged.}
\vspace{-0.05 in}
\label{tab_aug_valid}
\centering
\begin{tabular}{cccllllll}
\hline
\toprule
\multirow{2}{*}{Back-end}  & \multirow{2}{*}{Train Set} & \multirow{2}{*}{\begin{tabular}[c]{@{}c@{}}CKPT \\ Avg.\end{tabular}} & \multicolumn{3}{c}{w/ Aug.} & \multicolumn{3}{c}{w/o Aug.} \\ \cmidrule(r){4-6} \cmidrule(r){7-9}
& & & 21LA~\cite{ASVspoof2021} & 21DF~\cite{ASVspoof2021} & In-the-Wild~\cite{In_the_wild} & 21LA~\cite{ASVspoof2021} & 21DF~\cite{ASVspoof2021} & In-the-Wild~\cite{In_the_wild}  \\
\hline
\midrule
\multirow{3}{*}{Nes2Net-X} & \multirow{3}{*}{ASVspoof 19~\cite{ASVspoof2019}} & N/A & 1.63 (1.79) & 1.84 (2.03) & 5.56 (6.61) & 1.73 (1.95) & 1.65 (1.91) & 5.73 (6.83)  \\
&  & 3 & 1.70 (1.80) & 1.88 (1.98) & 5.15 (6.31) & 1.66 (1.87) & 1.54 (1.98) & 5.59 (6.90)\\
&  & 5 & 1.67 (1.78) & 1.80 (1.91) & 5.28 (6.31) & 1.88 (2.00) & 1.49 (1.78) & 5.52 (6.60)\\
\bottomrule
\hline
\end{tabular}
\vspace{-0.1 in}
\end{table*}

\subsection{{Empirical Runtime and Memory Analysis}}
{
Number of parameters and MMACs are widely adopted metrics for evaluating model efficiency. These platform-independent measures offer consistent and fair comparisons across different hardware. However, to better reflect the real-world deployment costs of back-end architectures, we additionally benchmark their training time, inference time, and peak GPU memory usage, as summarized in Table~\ref{tab_speed}.
}
\begin{table}[h]
\centering
\caption{{
Training and inference efficiency comparison across back-end models. The table reports the average (Avg.) training and inference time per batch in milliseconds (ms/batch), as well as peak GPU memory usage in megabytes (MB).}
}
\vspace{-0.05 in}
\setlength{\tabcolsep}{1.8mm}{
\begin{tabular}{lccc}
\hline
\toprule
\multirow{2}{*}{Back-end}  & \multicolumn{2}{c}{Avg. Time (ms/batch)↓}  & \multirow{2}{*}{Peak GPU Memory↓} \\ \cmidrule(r){2-3}
 & Training & Inference & (MB) \\ 
 \hline
\midrule
AASIST Light (C=24) & 27.0 & 7.8 &  1,327 \\
AASIST Standard(C=32) &  53.8 & 18.7 & 3,454 \\
AASIST Large(C=40) & 79.2 & 28.1 & 4,273 \\
AASIST XL(C=48) & 86.1 & 30.7 & 5,087  \\
AASIST XXL(C=56) & 100.9 & 37.4  &  5,905  \\ 
\hline
ResNet & 7.8 & 2.6 &  691  \\
Res2Net & 15.6 & 3.5 &  721 \\
ECAPA-TDNN (C=128) & 9.4 & 3.1 &  698\\ 
\hline
\textbf{Proposed Nes2Net} & 20.2 & 4.9 &  1,312 \\
\textbf{Proposed Nes2Net-X} & 29.1 & 9.2 &  2,231 \\

\bottomrule
\hline
\end{tabular}
}
\label{tab_speed}
\end{table}

All back-end models are evaluated under identical conditions: input features of 400 frames with 1024 dimensions, a batch size of 64, and execution on a dedicated NVIDIA H20 GPU. The first 10 batches are used for warm-up and excluded from the measurement, and the inference and training times are averaged over the subsequent 200 batches. Training time includes the forward, backward, and optimizer update steps.

The results show that AASIST models exhibit rapidly increasing runtime and memory consumption as the channel dimension $C$ grows. In contrast, our proposed Nes2Net achieves notably lower latency and memory usage. {Nes2Net-X} further improves performance in some settings by preserving more high-dimensional information, albeit at the cost of higher resource consumption.

Conventional models such as ResNet, Res2Net, and ECAPA-TDNN offer faster runtime and smaller memory footprints than our proposed method, but fall short in detection accuracy as shown in earlier experiments. Therefore, when selecting a back-end architecture, we believe both {Nes2Net} and {Nes2Net-X} offer flexible options: the former prioritizes efficiency, while the latter favors accuracy when computational resources permit. This underscores the importance of balancing performance and efficiency in real-world applications.

\subsection{Should We Use Augmentation During Validation?}
In all previous experiments, the datasets are split into three non-overlapping subsets: training, validation (or development), and test sets.
The validation set is used to select the best-performing checkpoints for final evaluation on the test set. The training set typically applies data augmentation to enhance model performance and generalization. However, the use of augmentation during validation remains inconsistent across prior studies. For instance, wav2vec 2.0-AASIST~\cite{hemlata_wav2vec2} applies the same augmentation strategy to both training and validation sets. In contrast, WavLM-AASIST~\cite{SVDD_i2r} does not use augmentation on the validation set, aligning with common practices in speaker verification research~\cite{RecXi, 10760244, ECAPA_TDNN}.

In this section, we compare these two approaches and report the results in Table~\ref{tab_aug_valid}. We observe that applying the same augmentation to the validation set as in the training set leads to worse performance on ASVspoof 2021 DF~\cite{ASVspoof2021}, but better results on In-the-Wild~\cite{In_the_wild}. When no augmentation is applied to the validation set, the opposite trend is observed.

From the outcome of the above study, we believe that in cases where robustness to certain variations (e.g., noise, compression, or distortions) is important, applying augmentation during validation provides insights into how well the model handles such conditions. As a result, the selected checkpoints from this approach may generalize better to these variations. Further investigation into this topic may yield deeper insights for future work.

\section{Conclusion}
In this work, we propose Nested Res2Net (Nes2Net) and its enhanced variant, Nes2Net-X, as lightweight and dimensionality reduction (DR) layer-free back-end architectures designed for speech anti-spoofing in the era of foundation models.
Unlike conventional approaches that rely on a DR layer to bridge the mismatch between high-dimensional features and downstream classifiers, our proposed architectures directly process these rich representations. This not only eliminates the computational and parameter overhead introduced by DR layers but also avoids information loss, enhancing overall system efficiency and robustness.

Nes2Net incorporates a novel nested multi-scale design that enables more effective feature extraction and deeper cross-channel interactions without increasing model complexity. The improved Nes2Net-X further strengthens representation learning by introducing learnable weighted feature fusion, offering adaptive control over the feature aggregation process.

We conduct extensive evaluations across five representative datasets: CtrSVDD, ASVspoof 2021, ASVspoof 5, PartialSpoof, and In-the-Wild, covering a wide range of singing voice deepfakes, fully spoofed speech, adversarial attacks, real-world deepfakes, and partially spoofed speech.
Across all scenarios, our models achieve SOTA performance, demonstrating superior generalization, compactness, and resilience under unseen and challenging conditions. 

In summary, Nes2Net and Nes2Net-X offer a general-purpose, resource-efficient back-end for foundation model-based speech anti-spoofing, providing a practical yet powerful alternative to DR-dependent designs. To facilitate future research and applications, we make all source code and pre-trained models publicly available.

\balance
\bibliographystyle{IEEEtran}
\bibliography{main}

\vfill

\end{document}